# Latest results from the IceCube Experiment.

Anthony M Brown[1], for the IceCube Collaboration [2]

[1]Department of Physics and Astronomy, University of Canterbury,
Private Bag 4800, Christchurch, New Zealand
anthony.brown@canterbury.ac.nz

[2]IceCube Collaboration, http://icecube.wisc.edu

*Abstract Summary*

*The IceCube Collaboration is currently building a neutrino detector at the South Pole to observe high energy neutrinos from a variety of astrophysical sources. In this paper we review the current status of the IceCube experiment, highlighting some of the results obtained so far.*

## I. Introduction

Neutrinos afford us a unique view of the Universe. Due to their small interaction cross-section and neutral charge, neutrinos are able to escape, undeflected, from some of the most dense regions of the Universe. As such, neutrinos offer us a unique probe of the most extreme processes occurring in cosmic sources.

The production of high energy neutrinos can occur via the decay of charged mesons, with the mesons being produced through the interaction of accelerated hadrons with ambient matter or radiation fields. In an astrophysical scenario, the production of high energy neutrinos is believed to occur in numerous sources where hadronic, or Cosmic Ray, acceleration is taking place. These sources include Active Galactic Nuclei, supernova remnants or microquasars [1,2,3]. Therefore, neutrino astronomy plays an important role, within the context of a multi-messenger approach, in conclusively proving the origin of Cosmic Ray particles.

Searching for neutrino signatures from Dark Matter (DM) annihilation is another key goal of the IceCube neutrino telescope. Through elastic scattering, weakly interacting massive particles (WIMPs), relics from the Big Bang, are believed to accumulate at the centre of massive celestial objects such as the Sun. This over density of WIMPs results in an increased self-annihilation rate of these primordial particles, producing a measurable high energy neutrino flux in the process. Observing the presence, or indeed absence, of a neutrino flux allows us to constrain the contribution of WIMPs to the universal DM population.

With a view to pursuing these scientific goals, among others, the IceCube collaboration is currently building the world's largest neutrino telescope in the clear glacial ice of Antarctica, at the geographic South Pole. Due to be completed in early 2011, the final detector will contain 5160 optical sensors deployed on 86 strings in 1 cubic kilometer of ice (see Figure 1).

The physical dimensions of IceCube have been optimized to detect all flavours of neutrinos ($\nu_e$, $\nu_\mu$ and $\nu_\tau$), over a wide range of energies, from ~20 GeV to beyond $10^9$ GeV, with unprecedented energy and angular resolution. Coupling these important detector characteristics with the sheer size of the detector allows IceCube to place some of the most restrictive limits to date, on a wide variety of astro-particle models.

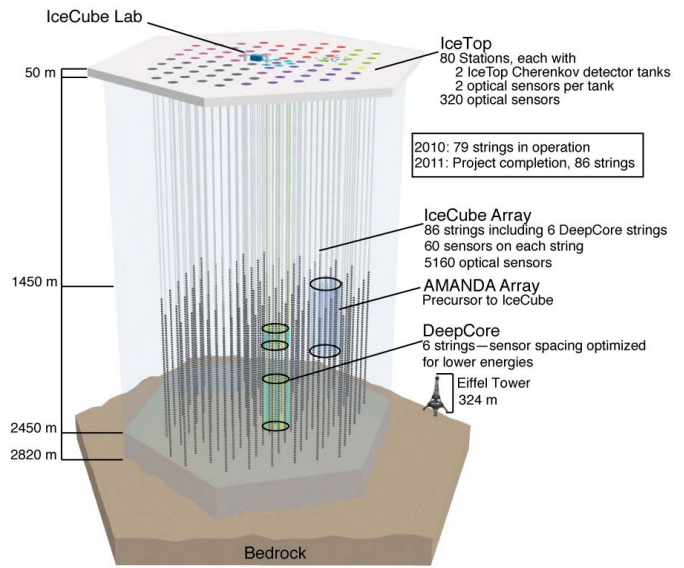

FIGURE 1: A schematic of the icecube detector.

After reviewing the detection technique and performance of the IceCube neutrino telescope, we will discuss some of the recent results of the IceCube experiment including the strongest muon flux limits from WIMP annihilation set to date, the results of our point source search and the first detection of Cosmic Ray anisotropy in the Southern sky.

## II. The IceCube Detector.

The IceCube neutrino telescope currently consists of 79 detector strings, each approximately 2.5 kilometers in length, covering a footprint of approximately 0.9 km$^2$. The physical dimensions of IceCube have been optimized for the detection of upward going muons and electromagnetic cascades over a wide range of energies. As such, each detector string supports 60 optical sensors equally spaced between depths of 1450 and

2450 meters. The vertical spacing between each optical sensor, is 17 meters, with approximately 125 meters between individual detector strings. Each optical sensor, referred to as a digital optical module, or DOM, consists of a pressure resistant glass sphere housing a 25 cm photomultiplier tube, LEDs for calibration purposes and two types of waveform digitisers (see Figure 2).

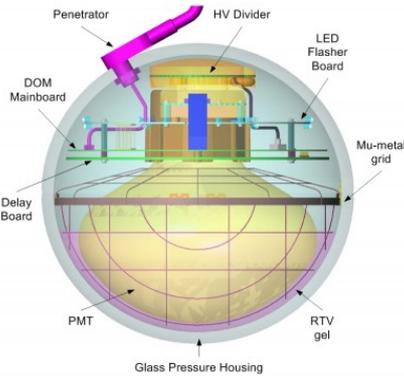

FIGURE 2: A SCHEMATIC OF AN ICECUBE DOM.

Complementary to these detector strings, IceCube also possesses a surface array and a densely instrumented core to extend the scientific capabilities of the detector. Referred to as IceTop, the surface array consists of 320 DOMs housed in 160 ice tanks, with 2 tanks located at the top of each detector string. The densely instrumented core detector, DeepCore, consists of 6 additional detector strings, with the DOMs positioned closer together, allowing us to lower the energy threshold of the detector by detecting fainter cascade and muon events.

The detection principle of IceCube relies on the observation of Cherenkov photons emitted by relativistic charged leptons produced through neutrino interactions with the surrounding ice and rock-bed, using a 3 dimensional lattice of DOMs. The lepton's passage through IceCube is reconstructed using temporal and positional knowledge of the Cherenkov photon hits on this 3-D lattice.

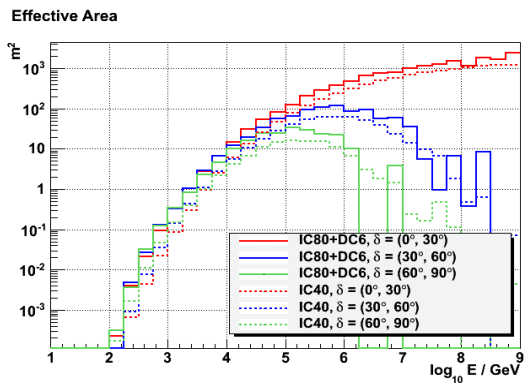

FIGURE 3: THE EFFECTIVE AREA OF ICECUBE AS A FUNCTION OF NEUTRINO ENERGY, FOR DIFFERENT DETECTOR CONFIGURATIONS.

With the DeepCore sub-detector, IceCube has an energy threshold of ~20 GeV for reconstructed neutrino events. Above this energy, IceCube's performance dramatically improves with increasing neutrino energy such that, at neutrino energies between 100 TeV - 10 PeV, IceCube has an effective area of ~100 $m^2$ - 1000 $m^2$ and sub-degree angular resolution. The performance of these important detector characteristics, as a function of neutrino energy, can be seen in Figures 3 and 4.

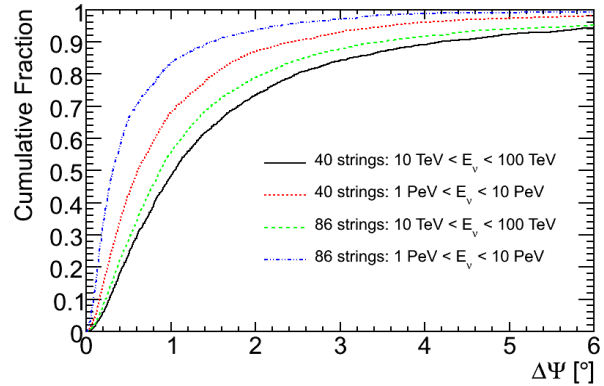

FIGURE 4: THE ANGULAR RESOLUTION FUNCTION OF ICECUBE FOR DIFFERENT ENERGY AND DETECTOR CONFIGURATIONS.

One method to confirm the angular resolution and absolute pointing of the IceCube detector is to use down-going muon events to observe the Moon's shadow on the Cosmic Ray (CR) flux. Observing the atmospheric muon flux allows IceCube to 'map' the CR flux around the moon, with a deficit in the observed muon flux being attributed to the Moon absorbing incident CRs. The results of such a study can be seen in Figure 5. With just 8 months worth of data from a 40 string IceCube detector, a ~5 sigma deficit in the CR flux in the direction of the Moon is clearly visible [6]. This deficit agrees with expectations and confirms both the angular resolution and absolute pointing of the IceCube detector.

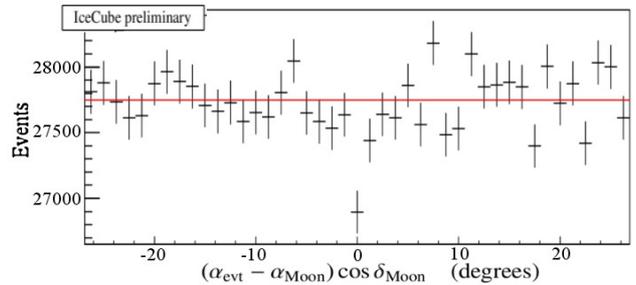

FIGURE 5: NUMBER OF EVENTS AS A FUNCTION OF 1.25 DEGREE DECLINATION BANDS CENTERED ON THE MOON USING 8 MONTHS OF IC40 DATA. A ~5 SIGMA DEFICIT IS SEEN IN THE CENTRAL BIN [6].

### III. RECENT RESULTS FROM ICECUBE

*A. Point Sources of Astrophysical Neutrinos*

One of the key scientific goals of IceCube is the search for point sources of astrophysical neutrinos. In pursuit of this goal, IceCube has conducted a point source search with the partially completed 40 string IceCube detector. The 40 string detector (IC40) was operational from April 2008 to May 2009, during which time ~ 3 x $10^{10}$ events were detected. The vast majority of these events were down-going atmospheric muon events. Two separate approaches have been used to remove the atmospheric muon background in the Northern and Southern skies. Since the IceCube detector is located at the South Pole, the Earth shields the detector from upward going atmospheric muons. Selecting up-going events allows us to define a final data set of the Northern sky sensitive to neutrinos in the TeV-PeV energy range. The point search can be extended to the down-going region on the basis that the expected source spectrum is harder ($E^{-2}$) than the atmospheric muon background spectrum ($E^{-3.7}$). Selecting high energy down-going events allows us to remove 5 orders of magnitude of atmospheric muon flux thus making a point source search of the Southern sky feasible.

After applying cuts to both energy and track reconstruction, the IC40 point source search was performed using an unbinned likelihood algorithm, that considers the data set to be a mixture of background and signal components [5]. Firstly we performed an all-sky search, scanning the whole sky in 0.1° x 0.1° bins. The results of the all-sky search can be seen in Figure 6, with the most significant deviation from the background being at 113.75° R.A and 15.15° Dec. However, after accounting for trial factors, this feature is statistically compatible with background fluctuations, since a spot with equal or higher significance was seen in 18% of the scrambled sky maps [10].

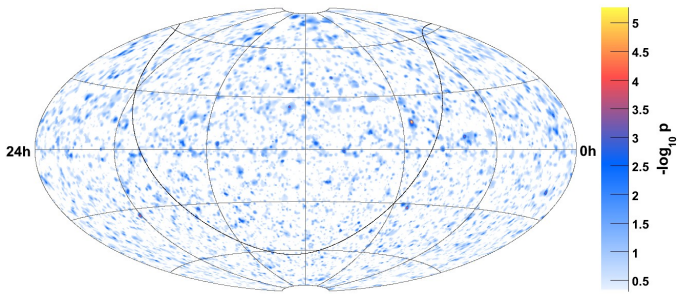

FIGURE 6: A PRE-TRIAL SIGNIFICANCE SKY MAP FOR THE ALL-SKY POINT SOURCE SEARCH USING IC40 DATA.

In order to avoid the large number of effective trials associated with scanning the entire sky, we also performed a search for a neutrino signal from a list of 39 candidate sources chosen a-priori. Considering these candidate sources allows us to improve the sensitivity since we are restricting our searches to a substantially smaller number of locations. For the 'candidate list' search, the most significant source was spatially coincident with PKS 1622-297. However, this feature is again consistent with background fluctuations. Assuming an $E^{-2}_\nu$ spectrum, neutrino flux limits for the candidate sources, as a function of declination angle, are shown in Figure 7.

### B. Indirect Dark Matter Search.

Within the minimal super-symmetric extension to the standard model, the neutralino is the lightest stable particle. With a mass in the GeV to TeV range, the neutralino interacts through gravity and the weak force alone; it therefore fulfills the requirements to be a WIMP cold dark matter candidate. In the context of neutrino telescopes, the search for DM is an indirect one. Due to elastic scattering, WIMPs may become gravitational bound to massive celestial objects, such as the Sun, resulting in an enhanced WIMP density and an increased self-annihilation rate. This increased WIMP self-annihilation rate can be observed as an excess high energy neutrino flux from these celestial objects.

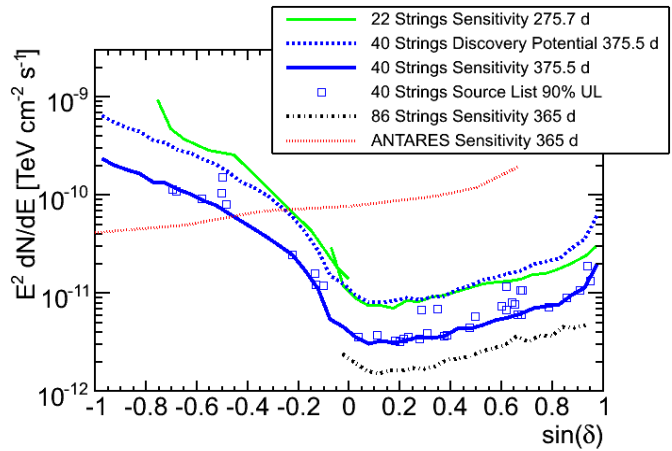

FIGURE 7: SENSITIVITY AND UPPER LIMITS (90% CL) PLOT FOR AN $E^{-2}$ SPECTRUM, AS A FUNCTION OF DECLINATION. THE SOLID BLUE LINE REPRESENTS THE UPPER LIMIT SET BY THE IC40 DATA, WITH THE GREEN LINE REPRESENTING THE UPPER LIMIT SET BY THE IC22 DATA. THE SQUARES REPRESENT THE UPPER LIMITS SET FOR THE 39 SOURCES SELECTED A-PRIORI.

To search for neutralino annihilation within the Sun, the partially completed 22 string (IC22) IceCube detector was used. To reduce the background, only up-going muon events were considered when the Sun was below the horizon. This results in 4.8 x $10^9$ events from 104.3 days of live time [4]. Applying a likelihood ratio method to this data set, the observed neutrino rate was compared to simulated neutrino flux levels predicted for a range of neutralino masses, via two extreme annihilation channels: hard (W boson) and soft (b quark) [4].

With no neutrino signal detected above the background, upper limits have been set following a unified Feldman-Cousins approach. These muon flux limits, for both annihilation channels, can be seen in Figure 8. These results are the most stringent limits to date on neutralino annihilation in the Sun.

### C. Cosmic Ray Anisotropy

While down-going atmospheric muons, originating from CR induced extensive air showers, are the dominate background for the IceCube detector, they can also be used to search for any large scale variations in the intensity of the CR flux. Such variations have been observed in the Northern hemisphere with both the Tibet and Milargro arrays [8,9], however to date, no such study has been completed for the Southern Hemisphere.

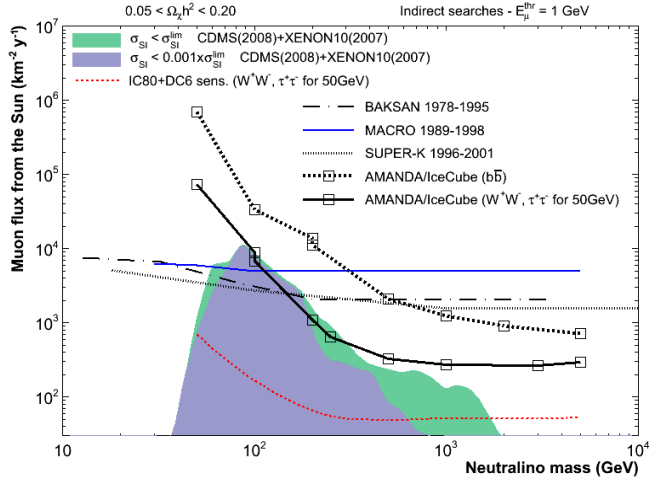

FIGURE 8: UPPER LIMITS AT THE 90% CL ON THE MUON FLUX FROM NEUTRALINO ANNIHILATION IN THE SUN, VIA BOTH SOFT AND HARD ANNIHILATION CHANNELS, AS A FUNCTION OF MASS. THE SHADED AREAS SHOWS THE MSSM MODEL NOT DISFAVOURED BY DIRECT SEARCHES.

To search for such features in the CR flux of the Southern sky, the partially completed 22 string detector (IC22) was again used. Considering all down-going muon events recorded between June 2007 and March 2008 and applying cuts on reconstruction and data quality resulted in a total of $4.3 \times 10^9$ atmospheric muon events with a median energy of 14 TeV per CR particle [7]. Using a likelihood based reconstruction algorithm, the 2-dimensional distribution of these muon events can be seen in Figure 9.

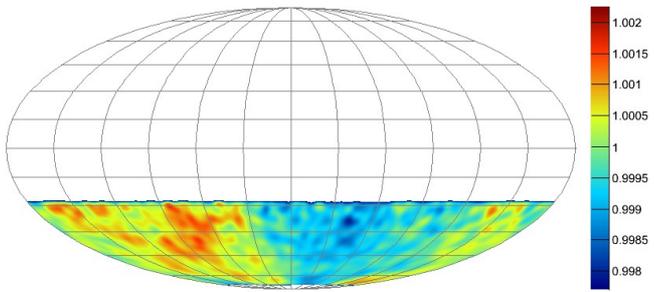

FIGURE 9: THE ICECUBE SKYMAP OF THE COSMIC RAY FLUX IN EQUATORIAL COORDINATES (DEC VS. RA). THE COLOR SCALE IS THE RELATIVE INTENSITY OF THE COSMIC RAY FLUX.

As seen in Figure 9, IceCube has found evidence for anisotropy CR flux intensity in the Southern sky [7]. A number of systematic checks have found the strength of this anisotropy to be independent of both the time of day and time of year. Furthermore, this anisotropy was found to persist at CR energies of up to 100 TeV, indicating that the Sun is not responsible for the observed variation in the large scale flux intensity variation. However, it should be noted that the magnitude of the anisotropy was found to have some energy dependence, following a decreasing trend with increasing CR energy.

IV. CONCLUSIONS

The IceCube neutrino telescope is close to completion, with 79 out of the planned 86 detector strings currently deployed. Nonetheless, throughout its construction, IceCube has been taking data. This data has allowed us to confirm both the effective area and angular resolution of IceCube over a wide range of energies.

Data taken during construction have also allowed IceCube to place some of the most stringent limits on astro-particle models to date. With the IC40 detector, no evidence was found for point sources of neutrinos. Assuming an $E^{-2}_\nu$ spectrum, upper limits on the neutrino flux have been set (see Figure 7). These flux limits are the most stringent set to date for high-energy neutrino point sources.

Data taken in the 22 string configuration have been used to probe neutralino annihilation in the Sun. While no neutrino signal has been observed above the background, we have been able to set the most restrictive limits to date, for muon flux rate from neutralino annihilation via 2 annihilation channels. The IC22 data set was also used to produce the first map of CR anisotropy for the Southern sky, complementing similar results from Northern hemisphere experiments (see Figure 9).

V. ACKNOWLEDGMENTS

IceCube acknowledge the support from the following agencies: U.S. National Science Foundation-Office of Polar Programs, U.S. National Science Foundation-Physics Division, University of Wisconsin Alumni Research Foundation, U.S. Department of Energy, and National Energy Research Scientific Computing Center, the Louisiana Optical Network Initiative (LONI) grid computing resources; National Science and Engineering Research Council of Canada; Swedish Research Council, Swedish Polar Research Secretariat, Swedish National Infrastructure for Computing (SNIC), and Knutand Alice Wallenberg Foundation, Sweden; German Ministry for Education and Research (BMBF), Deutsche Forschungsgemeinschaft (DFG), Research Department of Plasmas with Complex Interactions (Bochum), Germany; Fund for Scientific Research (FNRS-FWO), FWO Odysseus programme, Flanders Institute to encourage scientific and technological research in industry (IWT), Belgian Federal Science Policy Office (Belspo); Marsden Fund, New Zealand; Japan Society for Promotion of Science (JSPS); the Swiss National Science Foundation (SNSF), Switzerland; A.Gross acknowledges support by the EU Marie Curie OIF Program; J. P. Rodrigues acknowledges support by the Capes Foundation, Ministry of Education of Brazil.